\RequirePackage[2020-02-02]{latexrelease}
\documentclass[
 aapm,
 mph,
 amsmath,
 amssymb,
titlepage]{revtex4-1}

\makeatletter
\def\m@thit@licb@ld#1#2{\mbox{\boldmath$#1#2$}}
\def\mbit#1{\mathpalette\m@thit@licb@ld#1}

\usepackage{graphicx}
\usepackage{dcolumn}
\usepackage{bm}

\usepackage{float}
\makeatletter
\let\newfloat\newfloat@ltx
\makeatother

\usepackage{algorithm}
\usepackage{algorithmic}

\usepackage{siunitx}
\usepackage{xcolor}

\newcommand{\argmin}{\mathop{\rm arg~min}\limits}
\usepackage[mathlines]{lineno}
\usepackage{natbib}
\bibpunct[:]{(}{)}{;}{a}{,}{,}

\begin{document}
\title{Spatio-temporal reconstruction of substance dynamics using compressed sensing in multi-spectral magnetic resonance spectroscopic imaging}

\author{Utako Yamamoto$^{ab}$}
 \email{Corresponding author.\\ Room 408, Sogo-kenkyu 12th building, Yoshida-honmachi, Sakyo-ku, Kyoto-city, Kyoto, 606--8501, Japan}
 \email{E-mail addresses: yamamoto.utako.6a@kyoto-u.jp (U. Yamamoto)}
\author{Hirohiko Imai$^a$}
\author{Kei Sano$^a$}
\author{Masayuki Ohzeki$^{cdef}$}
\author{Tetsuya Matsuda$^a$}
\author{Toshiyuki Tanaka$^a$}

\affiliation{\footnotesize 
$^a$Department of Systems Science, Graduate School of Informatics, Kyoto University,  Kyoto, Japan,}

\affiliation{\footnotesize 
$^b$Research Fellow RPD, Japan Society for the Promotion of Science (JSPS)}

\affiliation{\footnotesize 
$^c$Graduate School of Information Sciences, Tohoku University, Sendai, Japan,}

\affiliation{\footnotesize 
$^d$International Research Frontier Initiative, Tokyo Institute of Technology, Tokyo, Japan,}

\affiliation{\footnotesize 
$^e$Department of Physics, Tokyo Institute of Technology, Tokyo, Japan,}

\affiliation{\footnotesize 
$^f$Sigma-i Co., Ltd., Tokyo, Japan,}



\maketitle
\noindent
\textbf{Abstract} The objective of our study is to observe dynamics of multiple substances in vivo 
with high temporal resolution from multi-spectral magnetic resonance spectroscopic imaging (MRSI) data. 
The multi-spectral MRSI can effectively separate spectral peaks of 
multiple substances and is useful to measure spatial distributions of substances. 
However it is difficult to measure time-varying substance distributions directly by ordinary full sampling 
because the measurement requires a significantly long time. 
In this study, we propose a novel method to reconstruct the spatio-temporal distributions of substances 
from randomly undersampled multi-spectral MRSI data on the basis of compressed sensing (CS) 
and the partially separable function model with base spectra of substances.
In our method, we have employed spatio-temporal sparsity and temporal smoothness of the substance distributions as prior knowledge to perform CS. 
By directly reconstructing the spatio-temporal distributions of the substances themselves 
without reconstructing the spectra, 
this method significantly reduces the amount of MRSI data required per single time frame.
We have formulated a regularized minimization problem for reconstruction and solved it by the alternating direction method of multipliers (ADMM). 
The effectiveness of our method has been evaluated using phantom data sets of glass tubes 
filled with glucose or lactate solution in increasing amounts over time 
and animal data sets of a tumor-bearing mouse to observe the metabolic dynamics 
involved in the Warburg effect in vivo.
The reconstructed results are consistent with the expected behaviors, 
showing that our method can reconstruct the spatio-temporal distribution of substances 
with a temporal resolution of four seconds which is extremely short time scale compared with that of full sampling.
Since this method utilizes only prior knowledge naturally assumed 
for the spatio-temporal distributions of substances 
and is independent of the number of the spectral and spatial dimensions or the acquisition sequence of MRSI, 
it is expected to contribute to revealing the underlying substance dynamics 
in MRSI data already acquired or to be acquired in the future.



\section{Introduction}
Magnetic resonance spectroscopic imaging (MRSI) has been recognized 
as a powerful tool to measure spatial distributions 
of chemical substances in vivo \citep{Posse2013MRSI}. 
It enables us to detect substances by identifying substance-specific spectral peaks. 
In multi-spectral MRSI, one can expect better separation of spectral peaks 
produced by different substances, 
and thus increasing the number of spectral dimensions of MRSI 
allows us to detect more kinds of substances \citep{Albert2001,Glunde2011MRS}. 
For example, 2D spectral MRSI acquired with a pulse sequence incorporating 
$^{1}$H-$^{13}$C heteronuclear multiple quantum coherence (HMQC) \citep{vanZijl1993gehmqc}
yields 2D spectra for H and C nuclei, 
which allow us to separate spectral peaks of lactate and fat, 
which are difficult with only the spectra for H nucleus. 
However, multi-spectral MRSI suffers from a prolonged acquisition time required to complete 
signal encodings for multiple spectral and spatial dimensions. 

The prolonged acquisition time prevents MRSI from detecting substances 
whose distributions vary in time. 
This is because the substance distributions will change 
before the signal encoding of multiple spectral and spatial dimensions is completed.
If the signal acquisition will be significantly accelerated, 
one will be able to observe spatio-temporal distributions of substances 
as a ``movie.'' 
Many acceleration methods for signal acquisition have been proposed \citep{Bogner2021}.
A method of combining multiband slice selection with consistent k-t-space EPSI has been developed for accelerated spectral imaging \citep{Schmidt2019}.
There is also a report of ultra-fast imaging of hyperpolarized biomolecules 
with 1D spectral MRSI using multi-echo balanced steady-state free precession sequences \citep{Muller2020}.
Although these methods have been successful in reducing the scanning time, 
acceleration is restricted by the performance of MR system 
and the design of acquisition pulse sequence. 
In order to circumvent such restrictions, 
an acceleration method which can be performed 
without special MR system or complicated acquisition pulse sequence 
has to be developed. 

The objective of this research is to reconstruct 
\emph{spatio-temporal dynamics} of substances in vivo 
from undersampled multi-spectral MRSI data acquired on time-varying samples 
on the basis of compressed sensing (CS) \citep{donoho2006compressed,lustig2007sparse}.
Undersampling is a way for acceleration 
without special MR system or complicated acquisition pulse sequence. 
It enables us to reduce the signal acquisition time 
by obtaining a small fraction of signals compared with full sampling. 
We employ a reconstruction method based on CS that directly 
estimates the spatio-temporal distributions of 
multiple substances 
by modeling the temporal variation of the substance distributions and 
using the base spectra of substances as prior knowledge.
It should be noted that in our method 
the substance distributions are not obtained via reconstructing the MRSI spectra, 
but by directly reconstructing the substance distributions themselves.

The CS provides a general framework for reconstruction from undersampled data 
with prior knowledge that variables to be estimated should be sparse.
In a previous study \citep{Santos-Diaz2019}, 1D MRSI spectrum was reconstructed with a 3-fold CS acceleration factor using sparsity of the spectrum in the wavelet-transformed domain.
The spatial dimensions in MRSI can provide various other types of 
sparsity besides the sparsity of MR spectra. 
For example, 
the sparsity about variations of spectra along spatial 
dimensions has been utilized in 1D spectral MRSI \citep{Saucedo2021} and 
2D spectral MRSI \citep{Iqbal2017}.
The sparsity of both of the 2D wavelet transform of spectra in the two spatial dimensions and the 3D total variation of spectra has been exploited in reconstruction of undersampled 1D spectral MRSI data \citep{Nassirpour2018}.
As a modeling of the temporal structure, 
\citet{Larson2011} reconstructed the substance dynamics with high temporal resolution 
by exploiting the sparsity of the wavelet transform of the spectrum 
in the time dimension. 
Their method used hyperpolarization of $^{13}$C 
to enhance the signal intensity of MRSI, 
and also constructed a pulse sequence suitable for the target substance, 
resulting in fast signal acquisition.

In addition, we have incorporated the partially separable function (PSF) model \citep{Liang2007,Haldar2010}, which is also known as the subspace model, into CS 
in order to achieve a temporal resolution that is much higher than the time scale of full sampling.
The PSF model exploits low-rankness of data 
in conjunction with low-rank matrix decomposition methods. 
By combining this idea with CS, 
MRSI data for the reconstruction of a 1D spectral-2D spacial MRSI image 
has been acquired in as short a time as 5 minutes \citep{Klauser2021}, 
and high-resolution imaging data has been created 
from the sparsely sampled and short-time acquired transients in mass spectrometry imaging \citep{Xie2022}.
The PSF model has also been employed to 
achieve the reconstruction of dynamic MRI images \citep{Feng2020,Djebra2022}.
A report by \citet{Li2021} has shown that high temporal and spatial resolution were achieved 
by using the PSF model and machine learning at the stage of data acquisition rather than reconstruction.
\citet{Lam2020} has reported an MRSI acceleration method 
that represents the high-dimensional spectroscopic imaging data 
as a union or superposition of subspaces by learning an efficient model 
to represent spectroscopic images using training data. 
Our method utilizes a pre-acquired spectrum as the base spectrum for each substance, 
but the base spectrum would also be obtained by learning parameters involved in the base spectrum, 
as in the previous study by \citet{Lam2020}. 
In other words, a future development of our method incorporating artificial intelligence is expected.

Our method does not utilize prior knowledge of the sparsity of the spectrum 
to perform CS, but only the sparsity that is supposed to exist in the substance distributions.
This is because our method is not aimed at reconstructing spectra, 
but at reconstructing the spatio-temporal distributions of substances.
In our method, the following two points were used as prior knowledge. 
One is that the substance distributions should have smoothness 
between adjacent time frames, 
since the substance does not appear or disappear suddenly, 
as is naturally assumed for the substance distributions,
and the other is about sparsity where the substance under consideration
exists only in a few spatial areas in the field of view 
and temporal intervals of the measurement.
We adopted only these natural points as the prior knowledge 
because we considered them to be the most reasonable and appropriate conditions 
for observing a variety of phenomena occurring in vivo.

To the best of our knowledge, 
no study has reported the method that reconstructs substance dynamic images per time frame with arbitrary length 
from multi-spectral MRSI 
by directly reconstructing the spatio-temporal distributions of multiple substances 
using the sparsity in the temporal variation of the substance distributions as a prior knowledge for CS.
Furthermore, the framework of our method is independent of the spectral and spatial dimensions of MRSI. 
In addition, the method can be applied to MRSI signals 
acquired with any MRSI pulse sequence, 
as long as they are randomly undersampled. 
The target substance is specified at the stage of reconstruction. 
The method is very systematic 
in that it is possible to reconstruct the spatio-temporal distributions 
of the target substances 
by simply performing the reconstruction with this method 
on MRSI data acquired in any dimensions and with any acquisition sequence.
Therefore, this method is expected to contribute to 
the effective utilization of both existing MRSI data and those to be acquired in the future.

\section{Theory}
This section describes a formulation to represent the reconstruction of the spatio-temporal distributions of substances as a minimization problem.
With reference to the PSF model, 
the ideal spatio-temporal spectrum is represented 
by a linear combination of the spectra of the substances.
The coefficients to be multiplied to the base spectra 
of the substances are the spatio-temporal distributions 
of the substances 
which are the targets of the estimation. 
Prior knowledge regarding the spatio-temporal distributions of substances, 
in both spatial and temporal domains, 
is incorporated into the formulation 
of CS as regularization. 
These formulations directly connect the spatio-temporal distributions of substances to be estimated 
with the measured MRSI signals. 
The substance distribution corresponding to each time point of the time-series of the measured MRSI signal is estimated, 
taking account of the distributions at the time points before and after that time point.

Let $J$ be the number of types of substances to be considered.
Let $x(\bm{r},t;j)$ denote the spatio-temporal distribution of substance $j\in\{1,\ldots,J\}$
at spatial position $\bm{r}$ at time $t$,
which we wish to estimate.
The ideal spatio-temporal spectrum $\Theta(\bm{f},\bm{r},t)$ as a function of 
frequency $\bm{f}$ associated with the spectral measurements, 
spatial position $\bm{r}$, and time $t$ is then given by 
\begin{equation}
\label{eq:position_phi}
\Theta(\bm{f},\bm{r},t) = \sum_{j=1}^J\theta_\mathrm{B}(\bm{f};j)x(\bm{r},t;j),
\end{equation}
where $\theta_\mathrm{B}(\bm{f};j)$ denotes 
the substance-specific spectrum which would be obtained 
if substance $j$ were measured in isolation without spatial encoding,
which we call the base spectrum.
In other words, the spatio-temporal spectrum is represented as a linear
combination of base spectra with coefficients
given by spatio-temporal distributions of substances. 
We write this linear relation abstractly as 
\begin{equation}
\label{eq:theta}
\Theta = \Theta_\mathrm{B} \bm{x}.
\end{equation}

Relationship of the signal $y$ and the spectrum $\Theta$ in MRSI is modeled as 
the discrete Fourier transform in the spatial dimensions as well as in the 
spectral dimensions.
Let $R\subset\mathbb{R}^\mathcal{N}$ denote the spatial region of interest,
which is assumed to be a direct product of intervals,
such as a rectangle for $\mathcal{N}=2$,
and appropriately discretized into a Cartesian grid.
Let $Q\subset\mathbb{R}^\mathcal{M}$ denote the measured range of spectra,
which is also assumed to be a direct product of intervals
and appropriately discretized into a Cartesian grid.
When $\mathcal{M}=1$, $Q$ corresponds to
the range of detection times in the readout direction.
When $\mathcal{M}\geq2$, the additional dimensions of $Q$
consist of axes of evolution times. 
The MRSI signal to be acquired
from the spatio-temporal spectrum $\Theta(\bm{f},\bm{r},t)$
at time $t$ is modeled as 
\begin{equation}
y(\mbit{\tau},\bm{k},t) = \mathcal{F}[\Theta(\bm{f},\bm{r},t)]
\end{equation}
for $(\mbit{\tau},\bm{k})\in\mathcal{F}(Q)\times\mathcal{F}(R)$, 
where $\mathcal{F}$ denotes the discrete Fourier transform operator,  
and where $\mathcal{F}(Q)$ and $\mathcal{F}(R)$ denote the dual grids 
of $Q$ and $R$, respectively, 
on which the discrete Fourier coefficients are defined.

If one could acquire the MRSI signals $\{y(\mbit{\tau},\bm{k},t)\}$
at all grid points of $\mathcal{F}(Q)\times\mathcal{F}(R)$ at time $t$,
then simple application of inverse discrete Fourier transform
would give us the spatio-temporal spectrum $\Theta(\bm{f},\bm{r},t)$ at time $t$.
When the spatio-temporal spectrum $\Theta(\bm{f},\bm{r},t)$
changes over time, however, one cannot acquire
the full set of MRSI signals 
$\{y(\mbit{\tau},\bm{k},t):(\mbit{\tau},\bm{k})\in\mathcal{F}(Q)\times\mathcal{F}(R)\}$
for the spatio-temporal spectrum $\Theta(\bm{f},\bm{r},t)$ at time $t$,
simply because of the prolonged acquisition time of MRSI.
If the time scale of temporal changes of the spatio-temporal spectrum 
is not too short, one could consider aggregating MRSI signals 
acquired within a time interval whose length is short relative to 
the time scale, such as the sliding window methods \citep{dArcy2002} 
and the keyhole technique \citep{Vaals1993}. 

In this paper, we take an approach based on the Bayesian framework,
in which $\bm{x}$ is estimated on the basis of
a posterior probability of $\bm{x}$ given MRSI signals.
The posterior probability is given in terms of a prior
and a likelihood.
Let $P(\bm{x})$ be an appropriate prior distribution
for the spatio-temporal distributions
$\bm{x}(\bm{r},t)$, $(\bm{r},t)\in R\times T$, 
where $T\subset\mathbb{R}$ denotes the time interval
over which we are interested in the spatio-temporal distributions. 
A likelihood function quantifies, given a spatio-temporal spectrum, 
how likely a set of acquired MRSI signals is. 
Let $\mathcal{T}=\{s_n\in T:n=1,2,\ldots\}$ denote 
the set of time points at which one acquires MRSI signals. 
We assume that MRSI signals are acquired at time $t=s_n$
on a subset $\mathcal{D}_n$ of grid points
in $\mathcal{F}(Q)\times\mathcal{F}(R)$,
for $n=1,2,\ldots$,
and let $\bm{y}(s_n)=\{y(\mbit{\tau},\bm{k},s_n):(\mbit{\tau},\bm{k})\in\mathcal{D}_n\}$
be the set of MRSI signals acquired at time $t=s_n$.
Let $P(\bm{y}|\Theta,\mathcal{D})$ denote the likelihood function of $\bm{y}$ 
acquired on $\mathcal{D}\subset\mathcal{F}(Q)\times\mathcal{F}(R)$
given $\Theta$. 
Let $\bm{y}_*=\{\bm{y}(s_n):s_n\in\mathcal{T}\}$ 
be a collection of acquired MRSI signals on $\mathcal{T}$. 
On the basis of the Bayesian framework,
we can then construct a posterior distribution
of $\bm{x}$ given $\bm{y}_*$ as
\begin{equation}
\label{eq:posd}
  P(\bm{x}|\bm{y}_*)
  \propto P(\bm{x})\prod_n P(\bm{y}(s_n)|\Theta_\mathrm{B}\bm{x}(s_n),\mathcal{D}_n).
\end{equation}
The maximum-a-posteriori (MAP) estimation of the spatio-temporal
distribution $\bm{x}$ given $\bm{y}_*$
is then formulated as the minimization of the negative log posterior, as
\begin{equation}
\label{eq:nlp}
  \min_{\bm{x}}\left(-\log P(\bm{x})-\sum_n\log P(\bm{y}(s_n)|\Theta_\mathrm{B}\bm{x}(s_n),\mathcal{D}_n)\right).
\end{equation}

Given MRSI signals 
$\bm{y}=\{y(\mbit{\tau},\bm{k}):(\mbit{\tau},\bm{k})\in\mathcal{D}\subset\mathcal{F}(Q)\times\mathcal{F}(R)\}$,
we define the likelihood function $P(\bm{y}|\Theta,\mathcal{D})$ as
\begin{equation}
\label{eq:lf}
  P(\bm{y}|\Theta,\mathcal{D})\propto
  \exp\left[-\frac{1}{2\sigma^2}
    \|\bm{y}-\mathcal{F}(\Theta)\|_{\mathcal{D}}^2
    \right],
\end{equation}
where for $\boldsymbol{\zeta}=\{\zeta(\mbit{\tau},\bm{k})\in\mathbb{C}:(\mbit{\tau},\bm{k})\in\mathcal{D}\}$ we let 
\begin{equation}
  \|\boldsymbol{\zeta}\|_{\mathcal{D}}^2
  =\sum_{(\mbit{\tau},\bm{k})\in\mathcal{D}}|\zeta(\mbit{\tau},\bm{k})|^2.
\end{equation}
It amounts to assuming that the noises in MRSI signal acquisition
are independent zero-mean complex Gaussian with variance $2\sigma^2$.

The prior distribution $P(\bm{x})$ summarizes
our beliefs, prior to data acquisition, about
how plausible a spatio-temporal distribution $\bm{x}$ is.
First, one can argue that in MRSI experiments 
we are basically interested in substances
which are more or less localized in space, 
and accordingly, a spatio-temporal distribution 
is expected to be sparse in space.
Second, metabolism in vivo consists of gradual processes,
so that temporal changes of the spatio-temporal distributions
are expected to occur smoothly. 
A prior distribution that incorporates the above two factors
may be given by
\begin{equation}
\label{eq:prid}
  P(\bm{x})\propto
  \exp\left[-\lambda_{\mathrm{x}} \|\bm{x}\|_1
    -\lambda_{\mathrm{W1}}\|\mathcal{W}\bm{x}\|_1
    -\frac{1}{2}\lambda_{\mathrm{W2}}\|\mathcal{W}\bm{x}\|_2^2
    \right],
\end{equation}
where $\|\mathcal{X}\|_p$, $p=1,2$, is the $\ell_p$-norm of $\mathcal{X}$,
and where $\mathcal{W}$ is a time-difference operator, 
defined as $\mathcal{W}\bm{x}(t)=\bm{x}(t+\Delta)-\bm{x}(t)$
with time difference $\Delta>0$. 
The first term in the exponent corresponds to the $\ell_1$-norm regularization and it promotes sparsity of the spatio-temporal distributions, 
whereas the second and the third terms correspond to what is called
the elastic net regularization \citep{ZouHastie2005} on $\mathcal{W}\bm{x}$
and they encourage temporal smoothness
of the spatio-temporal distributions. 
The regularization parameters $\lambda_\mathrm{x},\lambda_{\mathrm{W1}},\lambda_{\mathrm{W2}}>0$
control the relative strengths of the three terms
$\|\bm{x}\|_1$, $\|\mathcal{W}\bm{x}\|_1$, and $\|\mathcal{W}\bm{x}\|_2^2$. 

Our formulation of estimating
the spatio-temporal distributions $\bm{x}$
on the basis of acquired MRSI signals $\bm{y}_*$ is therefore
given by the convex minimization problem
\begin{eqnarray}
\label{eq:tv_minimization}
  \min_{\bm{x}}
  \left(\frac{1}{2}\sum_n\|\bm{y}(s_n)-\mathcal{F}(\Theta_\mathrm{B}\bm{x}(s_n))\|_{\mathcal{D}_n}^2
  +\lambda_{\mathrm{x}} \|\bm{x}\|_1 \right. 
  \left.+\lambda_{\mathrm{W1}}\|\mathcal{W}\bm{x}\|_1
  +\frac{1}{2}\lambda_{\mathrm{W2}}\|\mathcal{W}\bm{x}\|_2^2  {} \right),
\end{eqnarray}
which has a form of a regularized least-squares regression, 
and where the variance parameter $\sigma^2$
has been absorbed into the regularization parameters
$\lambda_{\mathrm{x}}$, $\lambda_{\mathrm{W1}}$,
and $\lambda_{\mathrm{W2}}$. 
We reconstruct the spatio-temporal distributions of substances 
by solving this minimization problem.

\section{Material and methods}
\begin{figure*}[tb]
\includegraphics[width=14cm]{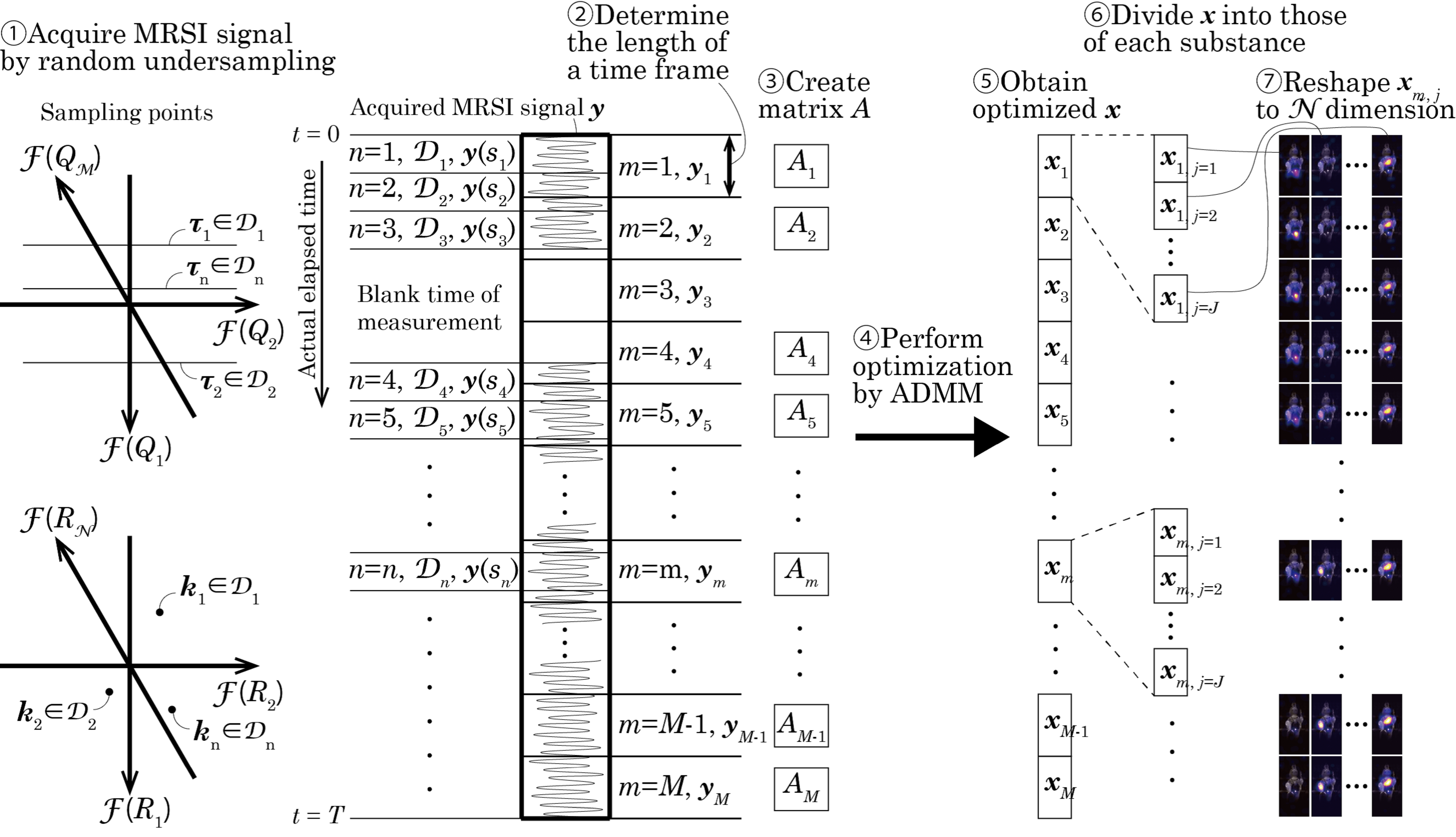}
\caption{A diagram of the proposed reconstruction method. 
The randomly undersampled MRSI signals are divided into time frames with arbitrary time width along the actual elapsed time 
and the spatio-temporal distributions of substances are assigned to the time frames.}
\label{fig:framework}
\end{figure*}

In this section, 
we describe the procedures of the phantom and the animal experiments, the imaging protocol in the data acquisition, 
how to set the time frame and construct the base spectrum in preparation for reconstruction, 
the algorithm of the optimization method, 
and the reconstruction setting.
In particular, the procedure described in subsection D, 
which divides the measured MRSI signal into time frames 
with arbitrary time width along the actual elapsed time 
and assigns the spatio-temporal distribution of substances 
to each time frame, 
is important for this reconstruction method.
A diagram of the proposed reconstruction method is shown in Fig.~\ref{fig:framework}.
\subsection{Phantom experiments}
We evaluated the performance of the reconstruction by our method using phantom experiments. 
MRSI data sets were acquired in three different settings using samples for which the spatial distributions of the substances and the time variations of their amount were known.
In Experiment 1, uniformly $^{13}$C-labeled D-glucose ([U-$^{13}$C$_6$]glucose) (Cambridge Isotope Laboratories, Inc., Andover, MA, USA) solution was instilled into a glass tube at a constant rate to increase the volume of the solution during the data collection. 
In Experiment 2, [U-$^{13}$C$_6$]glucose and [U-$^{13}$C$_3$]lactate 
solutions were instilled into the respective glass tubes.
In Experiment 3, in addition to the setting of Experiment 2, a live mouse was also placed between the glass tubes 
for data acquisition of fat. 
The treatment of the mouse followed the one described in the animal experiment section below.
The instilled volume of the solution was set to be increased along the slice direction of MR scans; 
since the MR scans were performed without slice selection, 
the increase in volume was measured as an increase in the spectral intensity of the corresponding voxels. 
In Experiment 1, a total of 400 $\mu$L of glucose solution was instilled at four different instillation rates of 53.3 (Experiment 1-1), 22.9 (Experiment 1-2), 11.1 (Experiment 1-3), and 5.9 (Experiment 1-4) $\mu$L/min, and the time required to instill the entire solution was 7.5, 17.5, 36.0, and 68.0 min, respectively. 
One tube was used to inject the glucose solution, 
and the four experiments with different settings in Experiment 1 were performed separately. 
In Experiments 2 and 3, a total of 400 $\mu$L of glucose solution and a total of 400 $\mu$L of lactate solution 
were instilled at rates of 22.9 $\mu$L/min for 17.5 min and 5.9 $\mu$L/min for 68.0 min, respectively. 
The instillation of both solutions into each of the two glass tubes was started simultaneously. 
The data acquisition started just prior to the beginning of instillation 
and continued until a set of 1,024 MRSI data were collected over 68 minutes.

\subsection{Animal experiment}
We furthermore evaluated our reconstruction method using animal experimental data in Experiment 4. 
In the experiment a tumor-bearing mouse injected with glucose solution was examined to observe metabolic processes related to the phenomenon called the Warburg effect \citep{Hsu2008WarburgEffect,Heiden2009WarburgEffect}. 
The Warburg effect refers to the phenomenon occurring in tumor cells which, unlike cells in normal tissues, exhibit high glycolytic activity 
and tend to convert most glucose to lactate, 
so that the amount of lactate in our experimental setting is expected to
increase gradually (about several minutes to several hours \citep{DeBerardinis2007}) 
in tumor following the injection. 
However it is difficult to observe the dynamics via full-sampling MRSI, 
because the process occurs in short time scale 
relative to the time required for full sampling of MRSI signals. 

All animal procedures were conducted in accordance with guidelines of 
animal experimentation at Kyoto University, where the experiment was conducted. 
A male BALB/c nude mouse (CLEA Japan, Inc., Tokyo, Japan) 
was transplanted with murine colon carcinoma cells 
(Colon 26, 1 $\times$ 10$^6$ cells / 50 $\mu$L) 
into subepidermal tissues on the right back near the shoulder 
several weeks prior to the MR experiments. 
Two molar of [U-$^{13}$C$_6$]glucose solution was prepared 
for the injection of glucose. 
Before the MR scanning, a 27-gauge butterfly needle attached to a 1-mL plastic syringe through a tubing filled with 
the [U-$^{13}$C$_6$]glucose solution was inserted into the abdomen. 
Throughout MR scanning, 
the mouse was anesthetized with 1--2.5 \% isoflurane (Forane; Abbott Japan, Co., Ltd., Tokyo, Japan) 
in a 1.5 L/min air flow through a plastic mask.
The injection of glucose solution was started at 4 minutes and 47 seconds 
passed from the beginning of the MR scanning. 
The injection continued at a constant rate of 10 $\mu$L/s, 
yielding a total injection time of 30 s.

\subsection{Data acquisition}
We acquired 2D spectral and 2D spatial MRSI data.
The MRSI pulse sequence we employed incorporates a gradient enhanced $^{1}$H-$^{13}$C 
heteronuclear multiple quantum coherence (HMQC) based preparation \citep{vanZijl1993gehmqc}.
For acceleration of data acquisition, we just employed random undersampling,
and did not use any fast-imaging methods specific to target substances
or 
any special accelerating techniques in the pulse sequence, in order to demonstrate that our method works efficiently without them. 

For designing the sequence of undersampling points in conjunction 
with the CS-based reconstruction, 
we employed Sobol sequences \citep{Sobol196786,niederreiter1992random}, 
which are one of low-discrepancy sequences. 
A low-discrepancy sequence is a pseudo-random sequence 
with the property that its finite-length subsequences are 
close to uniformly distributed. 
In actual MR scanning, 
signals can be acquired rapidly enough along readout trajectories without undersampling.  
Since $^{1}$H-$^{13}$C HMQC uses 
$^1$H spectral direction as the readout axis in $\mathcal{F}(Q_2)$, 
in our experiment, the sequence of undersampling points was determined 
for the remaining three axes of $^{13}$C spectral dimension in $\mathcal{F}(Q_1)$ and two spatial dimensions 
in $\mathcal{F}(R_1)\times\mathcal{F}(R_2)$. 

It is a common practice 
in designing undersampling points in MRSI \citep{furuyama2012application} 
that regions with higher signal-to-noise ratio (SNR) are
sampled more frequently than those with lower SNR. 
As for the $^{13}$C dimension, 
the sampling for small-$\mbit{\tau}$ points and large-$\mbit{\tau}$ points 
are corresponding to the regions with higher and lower SNR, respectively. 
To achieve such non-uniform sampling,
we applied nonlinear transformation to the elements of Sobol sequences
in $^{13}$C dimension. 
The particular form of the nonlinear transformation we used is as follows: 
Let $N_{\mathrm{C}}$ be the number of grid points 
in the $^{13}$C dimension of $\mathcal{F}(Q_1)$. 
Given an element $\eta\in[0,1)$, in the $^{13}$C dimension, of a member in a Sobol sequence
and a parameter $\psi>0$, 
we determined the corresponding index $d\in\{1,\ldots,N_{\mathrm{C}}\}$ via 
the following nonlinear transformation: 
\begin{equation}
d=\left\lfloor\frac{\log[1-(1-\psi^{N_{\mathrm{C}}})\eta]}{\log\psi}\right\rfloor+1.
\end{equation}
If $\eta$ is uniformly distributed over $[0,1)$, 
then the probability distribution of $d$ satisfies 
$P(d)\propto\psi^d$. 
We chose the value $\psi=e^{-4/N_{\mathrm{C}}}$ via our preliminary experiments, 
and used it in the following experiments. 
As for the spatial dimensions, on the other hand, 
we did not employ non-uniform undersampling, 
as our preliminary phantom experiments showed 
that non-uniform undersampling did not improve performance.
It can be ascribed to the fact that 
the spatial resolution in our experiments was 
relatively low. 
When the spatial resolution is higher, 
one might consider non-uniform undersampling 
in the spatial dimensions as well, 
with higher sampling points near the origin of the $k$-space. 

A preclinical 7T MR system (Bruker BioSpin MRI GmbH, Ettlingen, Germany) 
and a double resonant $^{1}$H/$^{13}$C transmit-receive volume coil 
(Doty Scientific, Inc., Columbia, SC, USA) were used. 
The following acquisition parameters were used: 
TR/TE = 990.7/10.7 ms, FOV = $4\times 4$ cm$^{2}$ (Experiment 1) and $4\times 8$ cm$^{2}$ (Experiments 2, 3 , and 4), 
coronal orientation without slice selection, 
numbers of points = 32 ($^{13}$C) and 256 ($^{1}$H), 
bandwidths = 2000 Hz ($^{1}$H) and 8000 Hz ($^{13}$C), 
imaging matrix size = $8\times 8$ (Experiment 1) and $8\times 16$ (Experiments 2, 3 , and 4), and 
four-step phase cycling. 
The MRSI acquisition took approximately 68 minutes 
for a sequence of undersampling points of length 1,024 with four-step phase cycling per point, 
which constitute what we call an MRSI acquisition session. 
We performed a single MRSI acquisition session in each of the phantom experiments (Experiments 1, 2, and 3), 
and repeated six MRSI acquisition sessions with the same sequence of undersampling points, 
yielding a set of 6,144 MRSI data in the animal experiment (Experiment 4). 
Exactly the same sequence of undersampling points determined by a Sobol sequence as described above was used for all MRSI acquisition sessions in Experiments 2, 3, and 4, whose settings share the same imaging matrix. 

In addition to the MRSI data, we acquired MRS data to construct the base spectra. 
The MRS data were obtained after the MRSI acquisition session (Experiments 1, 2, and 3), and every time between the consecutive MRSI acquisition sessions (Experiment 4),
with the same acquisition parameters as the MRSI acquisition except for the use of phase encoding gradients. 
In the MRS scan, we performed full sampling along the $^{13}$C spectral dimension, 
yielding full-sampled data in $\mathcal{F}(Q)$. 
The scanning time was 2.1 minutes. 
In Experiment 4, 
the total scanning time was 7.3 hours including the six MRSI and the five MRS acquisition sessions, 
as well as the preparation time between consecutive sessions.

\subsection{Preparation of reconstruction}
We discretize the whole elapsed time $T$ 
into $M$ equally-spaced time frames $\bar{\mathcal{T}}=\{t_m\in T:m=1,\ldots,M\}$, 
where $M$ is the total number of time frames in the entire time to be reconstructed. 
The spatio-temporal distributions $\bm{x}$ are then defined 
at these time frames, and let $\bm{x}_m=\bm{x}(t_m)$. 
For convenience, we assume that the set $\mathcal{T}$ of time points 
at which one acquires MRSI signals
is a subset of the set $\bar{\mathcal{T}}$ of time frames. 
If it is not the case, then one can interpolate $\bm{x}$ to evaluate 
the likelihood of the acquired MRSI signals at time points in $\mathcal{T}$. 
The MRSI acquisition had some blank time during the intervals 
between the consecutive MRSI acquisition sessions 
to acquire MRS data and to prepare the next acquisition session in Experiment 4.
As a consequence, there are time frames in $\bar{\mathcal{T}}$
that are not in $\mathcal{T}$. 
We here define $D=\{m:t_m\in\mathcal{T}\}$ as an index set to indicate the acquisition time points.
Then $|D|$ means the number of time frames where MRSI data exists.
The MRSI signals for $m\in D$ are collectively represented as $\bm{y}_m=\bm{y}(t_m)$ for simplicity.
Assuming the time difference $\Delta$ in the definition of the time-difference operator $\mathcal{W}$
to be equal to the time frame interval, 
our minimization problem [\ref{eq:tv_minimization}] is discretized as 
\begin{align}
\label{eq:tv_minimization_t}
\min_{\bm{x}}\left( \sum_{m\in D} \left\{ 
\frac{1}{2}\left\|\bm{y}_m - A_{m} \bm{x}_m \right\|_2^2 
+ \lambda_{\mathrm{x}} \left\| \bm{x}_m \right\|_1 \right\} \right. 
\left.+ \sum_{m=1}^{M-1} \left\{ \lambda_{\mathrm{W1}} \left\|\bm{x}_{m+1} - \bm{x}_{m} \right\|_1 
+ \frac{1}{2}\lambda_{\mathrm{W2}} \left\|\bm{x}_{m+1} - \bm{x}_{m} \right\|_2^2  \right\} \right), 
\end{align} 
where $A_m=U_mF\Theta_B$, 
where $U_m$ is the operator to extract elements corresponding to the undersampling points $(\mbit{\tau},\bm{k})\in\mathcal{D}$ at time $t_m$, and 
where $F$ is the inverse discrete Fourier transform matrix. 
Regularization with $\left\| \bm{x} \right\|_1$ 
is introduced only for time frames in $\mathcal{T}$:
If the regularization were introduced for time frames
in $\bar{\mathcal{T}}\backslash\mathcal{T}$ as well,
then $\bm{x}_m$ for $m\not\in D$ would tend to zero,
which is certainly undesirable. 

We prepared the base spectra $\Theta_B$ for target substances. 
We observed glucose in Experiment 1, glucose and lactate in Experiment 2, and fat as well as glucose and lactate in Experiments 3 and 4.
The natural abundance of $^{13}$C of approximately $1.1$ \%
yields significant MRSI signals from fat independent of
injection of $^{13}$C-labeled glucose,
which hampers accurate estimation of the amount of glucose and lactate. 
Distribution of fat is expected to be constant in time during the MRSI data acquisition, although we did not utilize the fact as prior knowledge of our reconstruction. 
In the 2D MRS spectrum of $^{1}$H-$^{13}$C HMQC, 
the spectral peak corresponding to fat is separable from those of glucose and lactate. 
We extracted a 2D spectrum corresponding to each of the three substances
from the spectra obtained by MRS measurements.
In the phantom experiments we used the data from the MRS acquisition session 
after the MRSI acquisition session. 
In the animal experiment, we employed the data from the second MRS acquisition session,
in which one can expect relatively large amounts of glucose and lactate
existing in the body of the mouse, yielding reliable base spectra for glucose and lactate.

\subsection{Optimization methods}
The optimization problem~\eqref{eq:tv_minimization_t} contains a large number of 
variables to be optimized.
We employed an optimization algorithm by \citet{Wahlberg2012},
which is based on the alternating direction method of multipliers (ADMM), 
to solve the problem. 
A basic idea behind the algorithm is to perform variable splitting, 
that is, introduction of auxiliary variables along with equality constraints, 
in order to make the resulting problems separable into smaller-sized subproblems, 
thereby allowing us to solve them efficiently. 
We show a pseudocode of our algorithm in \textbf{Algorithm~\ref{RSTpar}}, 
and describe it in the following. 
It should be noted that we do not claim that the particular form of the algorithm 
described below is optimal in any sense: It nevertheless demonstrates 
that the ADMM-based approach makes the large-scale optimization problem 
computationally feasible. 

We rewrite the minimization problem~\eqref{eq:tv_minimization_t}
via variable splitting as 
\begin{equation}
\begin{split}
\label{eq:f_min}
&\min_{\bm{x},\bm{h}} f(\bm{x},\bm{h})\\
&\mathrm{s.t.} \;\;(\bm{x},\bm{h}) \in C, 
\end{split}
\end{equation} 
where the objective function $f$ and the constraint set $C$ are
defined as 
\begin{align}
\label{eq:objective_constraint1}
&f(\bm{x},\bm{h}) = \sum_{m\in D} \left\{\Phi_m(\bm{x}_m) + \Xi_m(\bm{x}_m) \right\} + \sum_{m=1}^{M-1} \Psi_m(\bm{h}_m),\\
&\Phi_m(\bm{x}_m) = \frac{1}{2}\left\|\bm{y}_m - A_m \bm{x}_m\right\|_2^2,\\
&\Xi_m(\bm{x}_m) = \lambda_{\mathrm{x}} \left\|\bm{x}_m \right\|_1,\\
&\Psi_m(\bm{h}_m) = \lambda_{\mathrm{W1}} \left\|\bm{h}_m \right\|_1 
+ \frac{1}{2} \lambda_{\mathrm{W2}} \left\|\bm{h}_m \right\|_2^2,\\
&C = \left\{ (\bm{x},\bm{h}) : \bm{h}_m = \bm{x}_{m+1} - \bm{x}_{m}, \;\;m=1,\ldots,M-1 \right\} ,
\end{align}
with variables $\bm{x}=\{\bm{x}_1,\ldots,\bm{x}_M\}$ and 
$\bm{h}=\{\bm{h}_1,\ldots,\bm{h}_{M-1}\}$,
$\bm{x}_1,\ldots,\bm{x}_M,\bm{h}_1,\ldots,\bm{h}_{M-1} \in \mathbb{R}^{\mathcal{N} \times J}$. 
The constraint $(\bm{x},\bm{h})\in C$ can alternatively be represented 
using an indicator function to further rewrite~\eqref{eq:f_min} as 
\begin{equation}
\begin{split}
\label{eq:f_I_min}
&\min_{\bm{x},\bm{h},\bm{z},\bm{s}} \;\; \left\{f(\bm{x},\bm{h}) + \mathcal{I}_C(\bm{z},\bm{s}) \right\}\\
&\mathrm{s.t.} \;\; \bm{x} = \bm{z}, \; \bm{h} = \bm{s}, 
\end{split}
\end{equation} 
with variables $\bm{x} = (\bm{x}_1,\ldots,\bm{x}_M)$, $\bm{h} = (\bm{h}_1,\ldots,\bm{h}_{M-1})$, $\bm{z} = (\bm{z}_1,\ldots,\bm{z}_M)$, and $\bm{s} = (\bm{s}_1,\ldots,\bm{s}_{M-1})$, 
where $\mathcal{I}_C(\bm{z},\bm{s})$ is the indicator function on the constraint set $C$, taking 0 if $(\bm{z},\bm{s})\in C$, and $\infty$ otherwise. 

Following the standard prescription of ADMM, we introduce 
the augmented Lagrangian for our problem~\eqref{eq:f_I_min}, as
\begin{eqnarray}
\label{eq:Lagrangian1}
\mathcal{L}_{\rho_1,\rho_2}(\bm{x},\bm{z},\bm{u},\bm{h},\bm{s},\mbit{\nu}) = f(\bm{x},\bm{h}) + \mathcal{I}_C(\bm{z},\bm{s}) 
+ \frac{\rho_1}{2}\left\|\bm{x} - \bm{z} + \bm{u}\right\|_2^2 
+ \frac{\rho_2}{2}\left\|\bm{h} - \bm{s} + \mbit{\nu}\right\|_2^2 , 
\end{eqnarray}
where $\bm{u}$ and $\mbit{\nu}$ are scaled dual variables associated with the constraints 
$\bm{x} = \bm{z}$ and $\bm{h} = \bm{s}$, respectively,
and where $\rho_1,\rho_2 > 0$ are penalty parameters.
Each iteration of ADMM consists of four steps,
which will be described in the following. 

The first step in the ADMM iteration is minimization of the augmented Lagrangian over $\bm{x}$. 
At iteration $k$, we solve the following minimization subproblem  
\begin{eqnarray}
\label{eq:x_minfunc}
\bm{x}_m^{k+1} := \argmin_{\bm{x}} \left\{ \Phi_m(\bm{x}) + \Xi_m(\bm{x}) 
+ \frac{\rho_1}{2}\left\|\bm{x} - \bm{z}_m^k + \bm{u}_m^k \right\|_2^2 \right\} 
,\quad m=1,\ldots,M. 
\end{eqnarray}
We again make use of variable splitting to handle the above subproblem as 
\begin{equation}
\label{eq:mim_problem2}
\begin{split}
&\min_{\bm{x}, \mbit{\alpha}} \left\{ \Phi(\bm{x}) + \Xi(\mbit{\alpha})  + \frac{\rho_1}{2}\left\|\bm{x} - \bm{z}_m^k + \bm{u}_m^k \right\|_2^2 
 \right\}\\
&\mathrm{s.t.} \;\;\bm{x} = \mbit{\alpha},
\end{split}
\end{equation}
with $\bm{z}_m^k$ and $\bm{u}_m^k$ given. 
The augmented Lagrangian for this subproblem is 
\begin{align}
\label{eq:Lagrangian2}
\mathcal{L}_{\rho_1,\mu}(\bm{x},\mbit{\alpha},\mbit{\beta};\bm{z}_m^k,\bm{u}_m^k) = \Phi(\bm{x}) + \Xi(\mbit{\alpha}) 
+ \frac{\rho_1}{2}\left\|\bm{x} - \bm{z}_m^k + \bm{u}_m^k\right\|_2^2 
+ \frac{\mu}{2}\left\|\bm{x} - \mbit{\alpha} + \mbit{\beta} \right\|_2^2 , 
\end{align}
where $\mbit{\beta}$ is a scaled dual variable associated with the constraint $\bm{x}=\mbit{\alpha}$, 
and where $\mu > 0$ is a penalty parameter. 
We then apply the ADMM update
to the augmented Lagrangian~\eqref{eq:Lagrangian2}, yielding
an inner loop of the algorithm: at iteration $k'$ perform the updates
\begin{eqnarray}
\label{eq:x_minfunc2}
\bm{x}_m^{k,k'+1} &:=& \argmin_{\bm{x}} \left\{ 
\frac{1}{2}\left\|\bm{y}_{m} - A_{m} \bm{x}\right\|_2^2 \right. 
+ \left.\frac{\rho_1}{2}\left\|\bm{x} - \bm{z}_m^k + \bm{u}_m^k \right\|_2^2 
+ \frac{\mu}{2}\left\|\bm{x} - \mbit{\alpha}_m^{k,k'} + \mbit{\beta}_m^{k,k'} \right\|_2^2 
\right\}, \\
\label{eq:alpha_minfunc2}
\mbit{\alpha}_m^{k,k'+1} &:=& \argmin_{\mbit{\alpha}} \left\{ 
\lambda_{\mathrm{x}} \left\|\mbit{\alpha} \right\|_1 
+ \frac{\mu}{2}\left\|\mbit{\alpha} - (\bm{x}_m^{k,k'+1} + \mbit{\beta}_m^{k,k'}) \right\|_2^2 
\right\}, \\
\label{eq:beta_update}
\mbit{\beta}_m^{k,k'+1} &:=& \mbit{\beta}_m^{k,k'} + (\bm{x}_m^{k,k'+1} - \mbit{\alpha}_m^{k,k'+1}). 
\end{eqnarray}
The minimization problems~\eqref{eq:x_minfunc2}
and \eqref{eq:alpha_minfunc2} are explicitly solved, 
and the update in the inner loop of the algorithm is consequently derived as 
\begin{align}
\label{eq:x_update}
\bm{x}_m^{k,k'+1} &:= \left\{ \mathrm{Real}(A_m^{\mathrm{*tr}} A_m) + (\rho_1 + \mu)I \right\}^{-1} 
\left\{ \mathrm{Real}(A_m^{\mathrm{*tr}} \bm{y}_m) + \rho_1 (\bm{z}_m^k - \bm{u}_m^k) + \mu (\mbit{\alpha}_m^{k,k'} - \mbit{\beta}_m^{k,k'}) 
 \right\}, \\
\label{eq:alpha_update}
\mbit{\alpha}_m^{k,k'+1} &:= \mathrm{SoftThr} \left(\bm{x}_m^{k,k'+1} + \mbit{\beta}_m^{k,k'} ;\; \lambda_{\mathrm{x}} / \mu \right),
\end{align}
as well as~\eqref{eq:beta_update}, 
where the complex conjugate transpose is denoted by a superscript $\mathrm{*tr}$. 
The soft-threshold function 
$\mathrm{SoftThr} (\xi;\iota) \equiv \mathrm{sign}(\xi) \max \left\{\left|\xi\right| - \iota, 0\right\}$ \citep{Wright2009} was applied to update $\mbit{\alpha}_m$.
The inner loop is to be iterated an enough number of times.
The values of $\bm{x}_m^{k,k'}$ after an enough number of iterations
in the inner loop are to be used as the solution $\bm{x}_m^{k+1}$
of the subproblem~\eqref{eq:x_minfunc} in the first step.

The second step is established by the minimization of the augmented Lagrangian over $\bm{h}$, which amounts to solving 
\begin{eqnarray}
\label{eq:r_minfunc}
\bm{h}_m^{k+1} &:=& \argmin_{\bm{h}} \left\{ \Psi_m(\bm{h}) 
+ \frac{\rho_2}{2}\left\|\bm{h} - \bm{s}_m^k + \mbit{\nu}_m^k \right\|_2^2 \right\}\nonumber\\
&=& \argmin_{\bm{h}} \left\{ \lambda_{\mathrm{W1}} \left\|\bm{h} \right\|_1 \right.
+ \left.\frac{\rho_2(1+\lambda_{\mathrm{W2}}/\rho_2)}{2}\left\|\bm{h} - \frac{\bm{s}_m^k - \mbit{\nu}_m^k}{1+\lambda_{\mathrm{W2}}/\rho_2} \right\|_2^2 \right\}, \quad m=1,\ldots,M-1.
\end{eqnarray}
This minimization is again explicitly solved and the solution is given by 
\begin{align}
\label{eq:r_update}
\bm{h}_m^{k+1} := \mathrm{SoftThr} \left(\frac{\bm{s}_m^k - \mbit{\nu}_m^k}{1+\lambda_{\mathrm{W2}}/\rho_2} ;\; \frac{\lambda_{\mathrm{W1}}}{\rho_2(1+\lambda_{\mathrm{W2}}/\rho_2)} \right).
\end{align}

In the third step, we calculate
\begin{equation}
  (\bm{z}^{k+1},\bm{s}^{k+1})
  =\Pi_C(\bm{x}^{k+1}+\bm{u}^k,\bm{h}^{k+1}+\mbit{\nu}^k),
\end{equation}
that is, the projection of $(\bm{x}^{k+1}+\bm{u}^k,\bm{h}^{k+1}+\mbit{\nu}^k)$
onto the constraint set $C$. 
It is formulated as the following minimization problem:
\begin{eqnarray}
(\bm{z}^{k+1},\bm{s}^{k+1})=\argmin_{(\bm{z},\bm{s}); \bm{s}=W\bm{z}}
\left\{\rho_1\|\bm{x}^{k+1}-\bm{z}+\bm{u}^k\|_2^2 \right. 
\left. + \rho_2\|\bm{h}^{k+1}-\bm{s}+\mbit{\nu}^k\|_2^2
\right\}.
\end{eqnarray}
The solution is explicitly given by 
\begin{align}
\label{eq:z_update}
&\bm{z}^{k+1} = (LL^{\mathrm{tr}})^{-1} \left\{\bm{x}^{k+1} + \bm{u}^k + \gamma W^{\mathrm{tr}} (\bm{h}^{k+1} + \mbit{\nu}^k)\right\}, \\
\label{eq:s_update}
&\bm{s}^{k+1} = W \bm{z}^{k+1},
\end{align}
where $\gamma = \rho_2 / \rho_1$, 
the superscript $\mathrm{tr}$ represents the transpose of a matrix, 
$W$ is the matrix calculating the time difference, 
and $L$ denotes the Choleskey factorization of $I + \gamma W^{\mathrm{tr}}W$. 
That $L$ is a band matrix allows efficient implementation,
which is illustrated in \textbf{Algorithm~\ref{RSTpar}}
following the formulation by \citet{Wahlberg2012}. 

As the last step of ADMM, 
we update the scaled dual variables. 
\begin{align}
\label{eq:u_update}
&\bm{u}_m^{k+1} := \bm{u}_m^{k} + (\bm{x}_m^{k+1} - \bm{z}_m^{k+1}), \;\;m=1,\ldots,M, \\
\label{eq:nu_update}
&\mbit{\nu}_m^{k+1} := \mbit{\nu}_m^{k} + (\bm{h}_m^{k+1} - \bm{s}_m^{k+1}), \;\;m=1,\ldots,M-1.
\end{align}

All the updating processes described so far are summarized in \textbf{Algorithm~\ref{RSTpar}}. 
We implemented the algorithm on Matlab (ver.~R2017b, MathWorks, Inc., Natick, MA, USA).

\begin{algorithm*}[t]
\caption{Optimization procedure for our problem.}
\label{RSTpar}
\begin{algorithmic}
\REQUIRE $\bm{y}, A, L$;\quad 
$\lambda_{\mathrm{x}}, \lambda_{\mathrm{W1}}, \lambda_{\mathrm{W2}}$: Regularization parameters;\quad 
$\rho_1, \rho_2, \mu$: Penalty parameters;\quad 
$\gamma = \rho_2 / \rho_1$
\STATE $\bm{x}_m \leftarrow \mathrm{Real}(A_m^{\mathrm{*tr}} \bm{y}_m)$;\quad 
$\bm{z}_m \leftarrow \bm{x}_m$;\quad 
$\mbit{\alpha}_m, \mbit{\beta}_m, \bm{u}_m, \bm{h}_m, \mbit{\omega}_m, \bm{q}_m, \bm{s}_m, \bm{g}_m, \bm{b}_m, \mbit{\nu}_m \leftarrow 0$ \COMMENT{Initialization}
\WHILE{$\bm{x}_m$ not converged}
\STATE \COMMENT{Step 1, Eqs.~\eqref{eq:beta_update}--\eqref{eq:alpha_update}} 
\quad ($m = 1,2,\ldots,M$)
\WHILE{$\bm{x}_m$ not converged}
\STATE $\bm{x}_m \leftarrow \left\{ \mathrm{Real}(A_m^{\mathrm{*tr}} A_m) + (\rho_1 + \mu)I \right\}^{-1} \left\{ \mathrm{Real}(A_m^{\mathrm{*tr}} \bm{y}_m) 
+ \rho_1 (\bm{z}_m - \bm{u}_m) + \mu (\mbit{\alpha}_m - \mbit{\beta}_m) 
 \right\}$
\STATE $\mbit{\alpha}_m \leftarrow \mathrm{SoftThr} (\bm{x}_m + \mbit{\beta}_m ;\; \lambda_{\mathrm{x}} / \mu)$ \quad ($m\in D$), 
\quad $\mbit{\alpha}_m \leftarrow \bm{x}_m + \mbit{\beta}_m$ \quad ($m\notin D$)
\STATE $\mbit{\beta}_m \leftarrow \mbit{\beta}_m + (\bm{x}_m - \mbit{\alpha}_m)$
\ENDWHILE
\STATE\COMMENT{Step 2, Eq.~\eqref{eq:r_update}}
\quad ($m = 1,2,\ldots,M-1$)
\STATE $\bm{h}_m \leftarrow \mathrm{SoftThr} \left(\frac{\mbit{s}_m - \mbit{\nu}_m}{1+\lambda_{\mathrm{W2}}/\rho_2} ;\; \frac{\lambda_{\mathrm{W1}}}{\rho_2(1+\lambda_{\mathrm{W2}}/\rho_2)}\right)$
\STATE \COMMENT{Step 3, Eqs.~\eqref{eq:z_update}--\eqref{eq:s_update}}
\STATE $\mbit{\omega}_m \leftarrow \bm{x}_m + \bm{u}_m$, 
\quad $\bm{q}_m \leftarrow \bm{h}_m + \mbit{\nu}_m$ \quad ($m = 1,2,\ldots,M$)
\STATE $\bm{b}_1 \leftarrow \mbit{\omega}_1 - \gamma \bm{q}_1$, 
\quad $\bm{b}_M \leftarrow \mbit{\omega}_M + \gamma \bm{q}_{M-1}$, 
\quad $\bm{b}_m \leftarrow \mbit{\omega}_m + \gamma (\bm{q}_{m-1}-\bm{q}_m)$ 
\quad ($m = 2,3,\ldots,M-1$)
\STATE $\bm{g}_1 \leftarrow ({1}/L_{1,1}) \times \bm{b}_1$, 
\quad $\bm{g}_m \leftarrow (1/L_{m,m}) \times (\bm{b}_m - L_{m,m-1}\bm{g}_{m-1})$ 
\quad ($m = 2,3,\ldots,M)$
\STATE $\bm{z}_M \leftarrow (1/L_{M,M}) \times \bm{g}_M$, 
\quad $\bm{z}_m \leftarrow (1/L_{m,m}) \times (\bm{g}_m - L_{m+1,m}\bm{z}_{m+1})$ 
\quad ($m = M-1,M-2,\ldots,1$)
\STATE $\bm{s}_m \leftarrow \bm{z}_{m+1} - \bm{z}_m$ \quad ($m = 1,2,\ldots,M-1$)
\STATE\COMMENT{Step 4, Eqs.~\eqref{eq:u_update}--\eqref{eq:nu_update}}
\STATE $\bm{u}_m \leftarrow \bm{u}_m + (\bm{x}_m - \bm{z}_m)$, 
\quad $\mbit{\nu}_m \leftarrow \mbit{\nu}_m + (\bm{h}_m - \bm{s}_m)$ \quad ($m = 1,2,\ldots,M$)
\ENDWHILE
\ENSURE $\bm{x}_m$
\end{algorithmic}
\end{algorithm*}

\subsection{Reconstruction setting}
The interval of time frames was set to four seconds. 
This interval corresponds to the acquisition time of MRSI data for a single undersampling point. 
The total number $M$ of time frames and the number of the time frames in $D$ having MRSI data were set as $M=|D|=1,024$ in the phantom experiments and $M=6,604$ and $|D|=6,144$ in the animal experiment, respectively.
In the phantom experiments, since we reconstructed the substance distributions from the data of a single acquisition session 
with no gaps during the total time of MRSI measurement, all time frames in $\bar{\mathcal{T}}$ have MRSI data.
Thus, in our settings, $M$ and $|D|$ are the same in the phantom experiments.
In the animal experiment, on the other hand, $M > |D|$ due to the multiple gaps between MRSI acquisition sessions.

The optimization for the reconstruction via the ADMM requires specification of the penalty parameters $\rho_1$, $\rho_2$, and $\mu$. 
In principle, the reconstructed result should not depend on the choice of the penalty parameters 
if the algorithm were iterated infinitely,
as our problem to be solved is a convex optimization problem. 
In practice, however, the choice of the penalty parameters
affects convergence speed of the algorithm and behavior in convergence to the solution. 
If one uses too small values for them,
the algorithm would require many iterations 
until convergence is achieved. 
With too large values, on the other hand,
the algorithm would tend to be unstable. 
We selected $\rho_1 = \mu = 10^{-3}$ and $\rho_2 = 10^{-1}$
on the basis of our preliminary experiments, 
which empirically allowed fast convergence of the algorithm. 

We set the total number of iterations to solve 
the optimization problem~\eqref{eq:f_I_min} to 1,000, 
and performed the optimization of the subproblem~\eqref{eq:x_minfunc} 
in Step 1 of \textbf{Algorithm~\ref{RSTpar}} with two iterations 
of Eqs.~\eqref{eq:beta_update}--\eqref{eq:alpha_update}. 
These were determined on the basis of the following observation. 

\begin{figure}[tb]
\includegraphics[width=6cm]{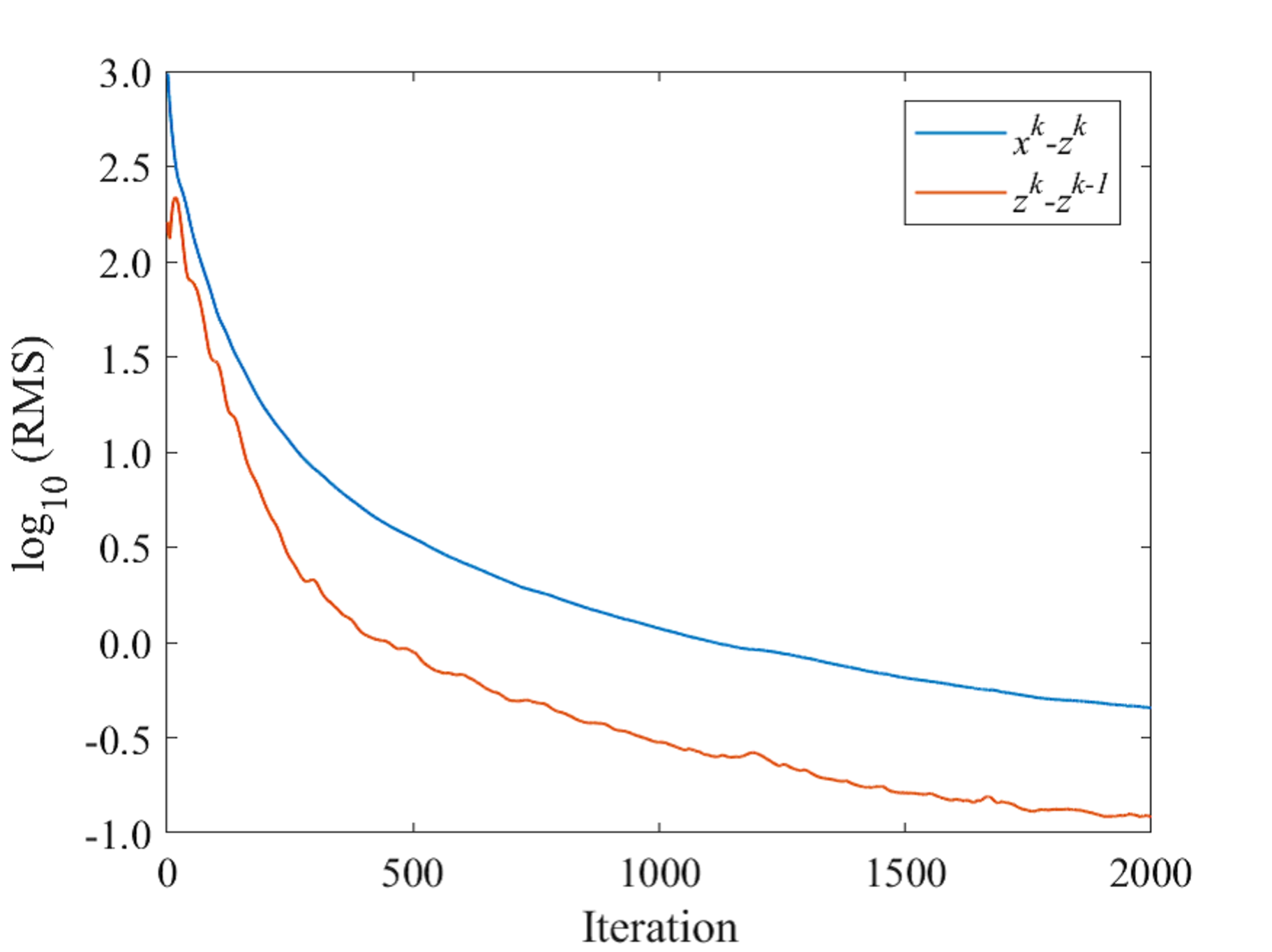}
\caption{The root mean squares (RMSs) of $\bm{x}^k-\bm{z}^k$ 
and $\bm{z}^k-\bm{z}^{k-1}$ at $k$th iteration. 
The vertical axis is in the logarithmic scale. 
Around the 500th iteration, the slope of the curves became more gradual.}
\label{fig:convergence}
\end{figure}

We show in Fig.~\ref{fig:convergence} how two residuals $\|\bm{x}^k-\bm{z}^k\|_2$ and $\|\bm{z}^k-\bm{z}^{k-1}\|_2$ changed in the execution of our algorithm. 
These two residuals were employed in \citet{Wahlberg2012} to assess convergence of optimization by ADMM. 
Since the values had a tendency to decrease slowly after around 500th iteration, 
we selected 1,000 iterations in our algorithm. 
As for the inner loop performing the minimization in Step 1, 
the first iteration brought us closer to the optimal solution, 
and after the second iteration, the value remained almost the same.

\section{Results}

\begin{figure*}[tb]
\centering
\includegraphics[width=13cm]{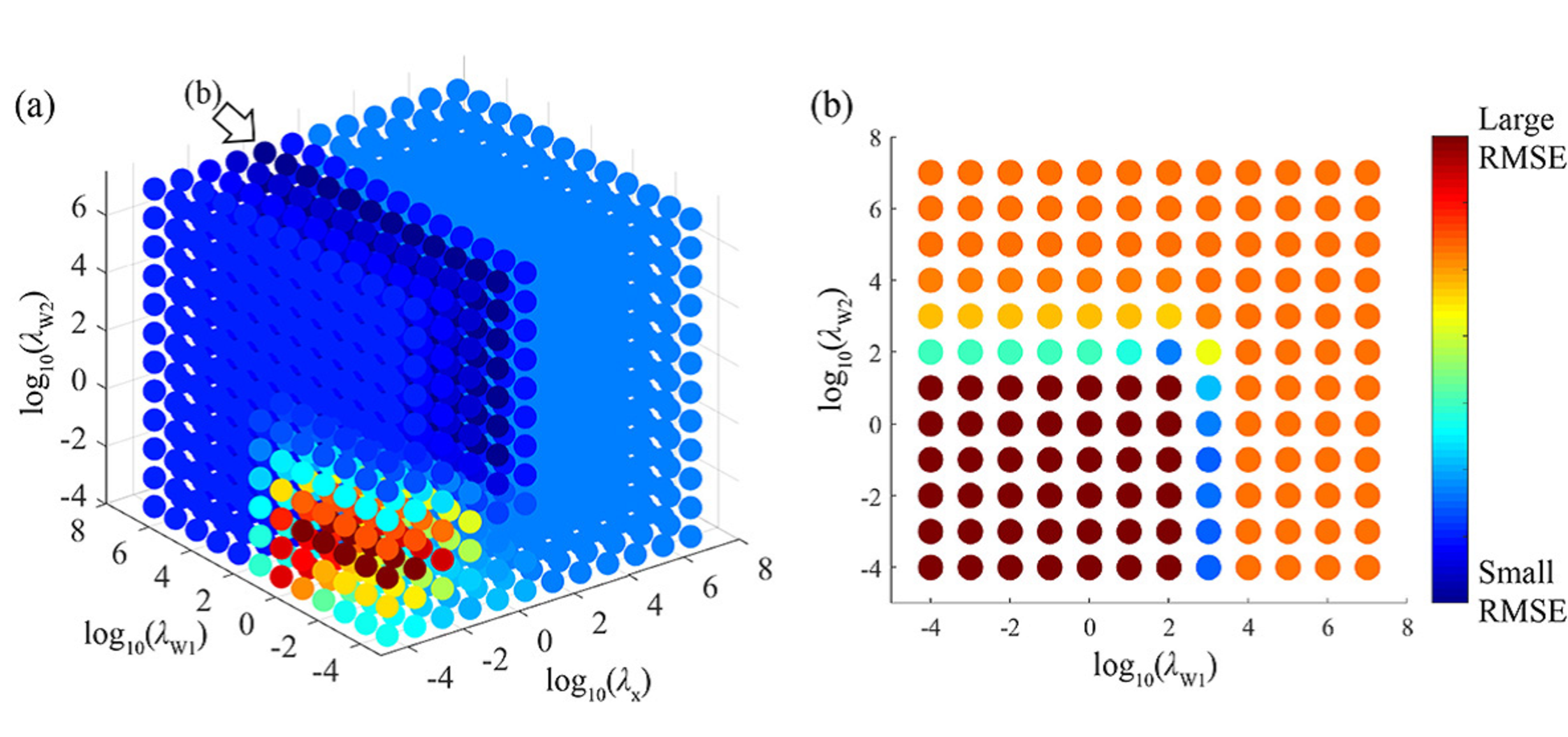}
\caption{The RMSE of the 2-fold CV for combinations of the regularization parameters in Experiment 4. 
(a) All combinations. (b) Combinations with $\lambda_{\mathrm{x}} = 10^0$.
Note that the same color range covers different value ranges in figures (a) and (b) 
to provide the visualization to identify large and small values.}
\label{fig:CV}
\end{figure*}

\begin{figure*}[tb]
\includegraphics[width=14cm]{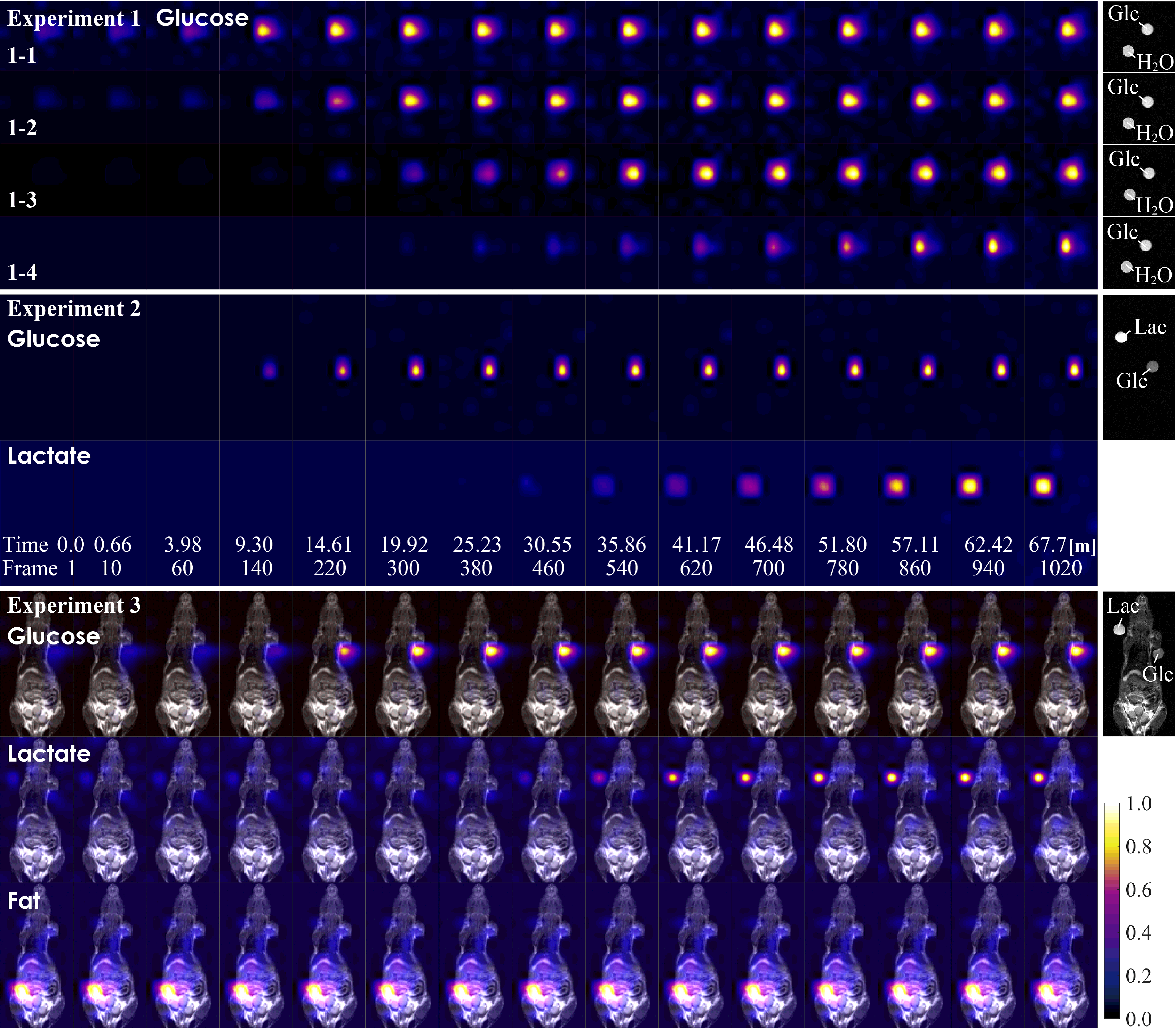}
\caption{Reconstructed spatio-temporal distributions of each substance in the phantom experiments (Experiments 1, 2, and 3). 
One column shows the time variation of one substance in one experiment.
Each figure shows a colored map of the spatial distribution of a substance in a single time frame.
In Experiment 3, 
images are displayed as colored maps overlaid on the $^1$H T$_2$-weighted images of mouse 
shown in gray scale for anatomical reference. 
For each substance, values of $\bm{x}$ were rescaled
so that the maximum value is equal to 1.
The "Frame" indicates the time frame index of the reconstruction 
and the "Time" indicates the elapsed time (in minute) from the beginning of the scanning. 
The images lined up vertically throughout all the experiments are the results of the same time frame.
The rightmost column shows the T2-weighted images taken immediately after the end of each experiment.}
\label{fig:dynamics_Phantom}
\end{figure*}

\begin{figure*}[tb]
\includegraphics[width=14cm]{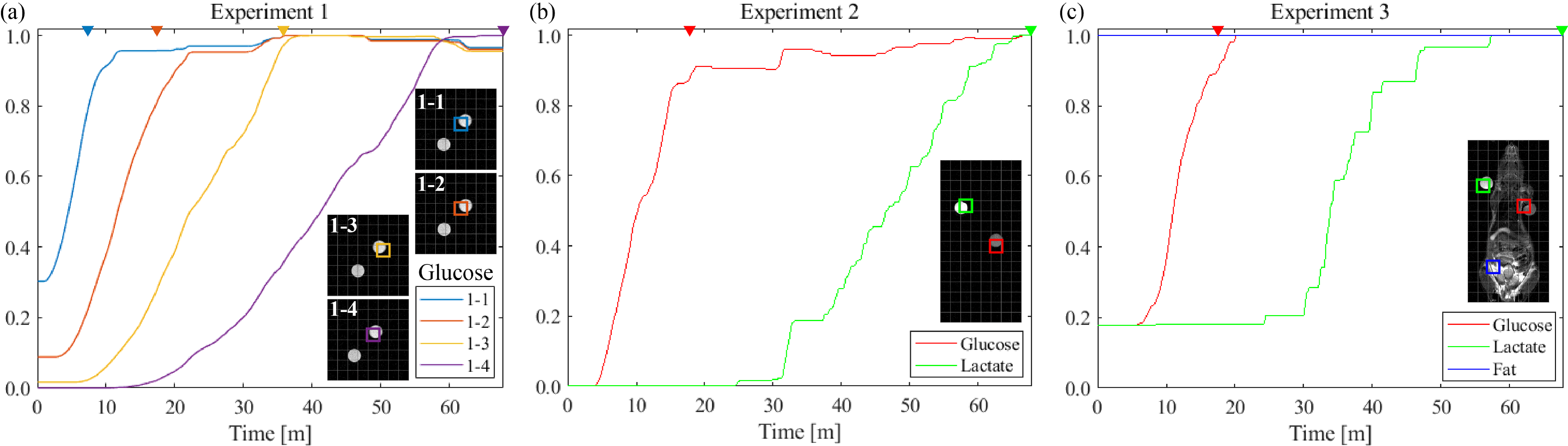}
\caption{Reconstructed temporal profiles of the amount of substances 
at spatial pixels with the largest value for each substance in the phantom experiments. 
The locations of the selected pixels are indicated by square markers in the T2-weighted image in each figure.
Horizontal axis represents the elapsed time in minute.
Vertical axis represents values of $\bm{x}$,
rescaled so that the maximum value is equal to 1 in each substance.}
\label{fig:temporal_Phantom}
\end{figure*}

\begin{figure*}[tb]
\includegraphics[width=14cm]{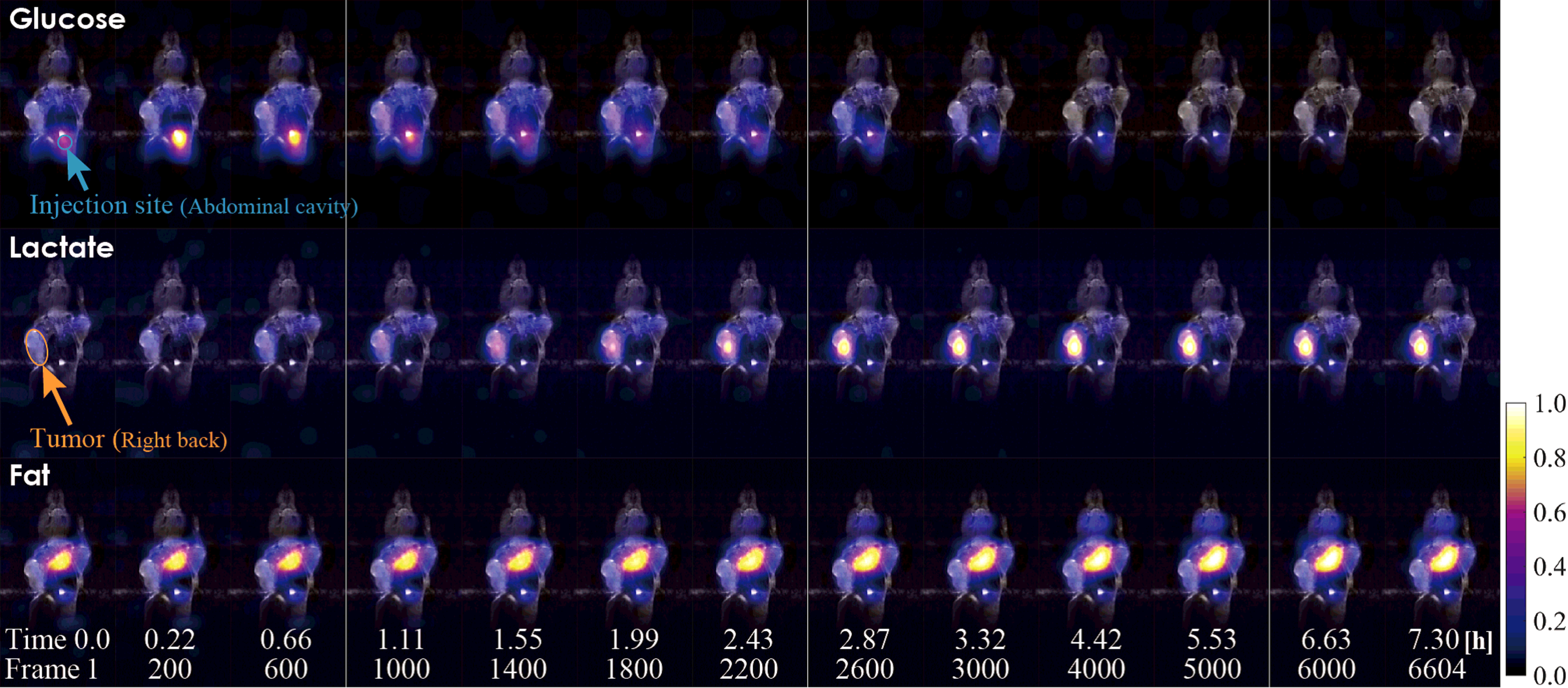}
\caption{Reconstructed spatio-temporal distributions of glucose (upper row), 
lactate (middle row), and fat (lower row) in the animal experiment (Experiment 4). 
Images are displayed as colored maps overlaid on the $^1$H T$_2$-weighted images. 
For each substance, values of $\bm{x}$ were rescaled
so that the maximum value is equal to 1.
Elapsed times from the beginning of the scanning (in hour)
and corresponding time frame indices are also shown at the bottom.}
\label{fig:dynamics_animal}
\end{figure*}

\begin{figure*}[tb]
\includegraphics[width=14cm]{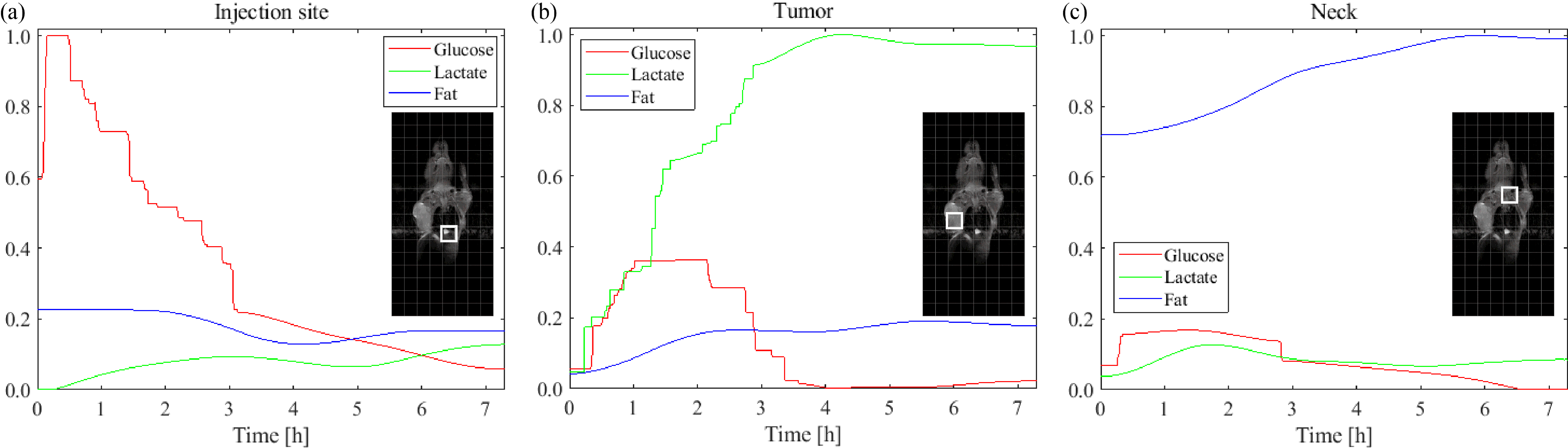}
\caption{Reconstructed temporal profiles of the amount of substances 
at spatial pixels corresponding to the injection site (a), the tumor (b), and the neck (c) in the animal experiment (Experiment 4). 
The pixels are indicated by white square markers in T$_2$-weighted anatomical images. 
Horizontal axis represents the elapsed time in hour.
Vertical axis represents values of $\bm{x}$,
rescaled so that the maximum value is equal to 1 in each substance.}
\label{fig:temporal_animal}
\end{figure*}

First, we determined the three regularization parameters 
$\lambda_\mathrm{x}$, $\lambda_\mathrm{W1}$, 
and $\lambda_\mathrm{W2}$ via 2-fold cross validation (CV) 
as follows: 
The MRSI data were divided into two subsets according to the order of the scanned readouts, 
with the odd-numbered readouts as subset 1 
and the even-numbered readouts as subset 2.
Estimation of $\bm{x}$ was performed with one of the subsets, 
and the quality of the estimated $\bm{x}$ was 
assessed with the other subset,
via calculating the values of $\bm{y}$ corresponding to the latter subset
from the estimated $\bm{x}$, and then 
evaluating the root mean square error (RMSE) of the calculated values of $\bm{y}$
against the readouts in the latter subset.
In other words, regarding each subset as a dataset
which has some missing readouts, 
we used it to estimate $\bm{x}$ for all time frames and 
calculated the values of $\bm{y}$ for the missing readouts using the estimated $\bm{x}$. 
The values of the parameters $\lambda_\mathrm{x}$, $\lambda_\mathrm{W1}$, and 
$\lambda_\mathrm{W2}$ were chosen from $\{10^{-4},10^{-3},10^{-2},\ldots,10^6,10^7\}$. 
We show the results of CV in Experiment 4 in Fig.~\ref{fig:CV}, where parameter combinations 
having high and low RMSEs are depicted in red and blue, respectively. 
The axes in Figs.~\ref{fig:CV} (a) and (b) are in the logarithmic scale. 
One observes in Fig.~\ref{fig:CV} (a) that the RMSE values were nearly 
unchanged when $\lambda_\mathrm{x}$ is large. 
Figure~\ref{fig:CV} (b) shows the results with $\lambda_\mathrm{x}=10^0$ 
in the $\lambda_\mathrm{W1}$-$\lambda_\mathrm{W2}$ plane, 
where the CV achieved smaller RMSEs for the most part.
We found minimizers of the RMSE values among all of the $12^3$ possible parameter combinations
and set them as the values of the regularization parameters: 
$(\lambda_{\mathrm{x}},\lambda_{\mathrm{W1}},\lambda_{\mathrm{W2}})$ as $(10^1,10^3,10^0)$ (Experiment 1), $(10^1,10^2,10^{-1})$ (Experiment 2), and  $(10^0,10^3,10^{-1})$ (Experiments 3 and 4). 

With the regularization parameters selected above, 
we performed the reconstruction using all the acquired MRSI data. 
Figures~\ref{fig:dynamics_Phantom} and \ref{fig:dynamics_animal} show snapshots of the reconstructed spatio-temporal distributions. 
For each time frame and for each substance, the dimension of $\bm{x}$ is $8\times8$ for Experiment 1 and $8\times16$ for Experiments 2, 3, and 4, 
corresponding to the spatial resolution of MRSI. 
For visualization, we post-processed the estimated $\bm{x}$ via 
first performing up-sampling of $\bm{x}$ with the bi-cubic interpolation, 
and then rescaling the values by setting the maximum value to unity for each substance. 
Each resulting image of $\bm{x}$ was overlaid on T$_2$-weighted MR image of the mouse in Experiments 3 and 4. 
Each image shows the ventral view of the mouse. 

Figures~\ref{fig:temporal_Phantom} and \ref{fig:temporal_animal} 
show the estimated temporal profiles of the amount of substances. 
The horizontal axis in Figures~\ref{fig:temporal_Phantom} and \ref{fig:temporal_animal} 
represents the elapsed time from the start of the MRSI measurement; 
the unit in Figure~\ref{fig:temporal_Phantom} is minute, and the unit in Figure~\ref{fig:temporal_animal} is hour.
In Figure~\ref{fig:temporal_Phantom}, 
the temporal variation of the estimated $\bm{x}$ of the spatial pixel 
(indicated by the square markers on the T2-weighted images in the figures) 
where the amount of each substance in each experiment took its maximum value 
is depicted.
It should be noted that the vertical axis represents 
values of substance amounts, rescaled so that
the maximum value of the reconstructed spatio-temporal distribution
of each substance is 1.
Each curve then represents relative temporal changes of the amount of the substance.
In Figure~\ref{fig:temporal_Phantom} (a), 
we can observe the increase in glucose instilled at different instillation rates. 
In Figures~\ref{fig:temporal_Phantom} (b) and (c), 
an increase in lactate was observed after the increase in glucose.
In Experiments 1-1--1-4, the instillation of glucose solution 
ended at 7.5, 17.5, 36.0, and 68.0 min, respectively. 
In Experiments 2 and 3, the instillation of glucose and lactate ended at 17.5 and 68 min, respectively. 
In Figure~\ref{fig:temporal_Phantom}, $\bigtriangledown$ indicates the time 
at which the instillation of each solution was finished.

Figures~\ref{fig:temporal_animal} (a)--(c) show temporal profiles of the three substances 
at the injection site, inside the tumor, 
and at the neck which is a lipid abundant region, respectively. 
Each of these positions is indicated in the inset of each figure with a square marker. 
The values of the substance amounts were rescaled
so that the maximum value of the reconstructed spatio-temporal
distribution of each substance is 1. 
The increase and decrease of glucose and lactate, and the constancy of fat were 
observed in Figures~\ref{fig:dynamics_animal} and \ref{fig:temporal_animal}. 
Glucose spread over the whole body in the early stage and decayed in the latter half. 
At the injection site, glucose increased quickly in the beginning, 
and decreased slowly after that. 
Inside the tumor, 
glucose increased more slowly than that at the injection site, 
and decreased in the latter half. 
Lactate increased following the rise of glucose 
and kept a high level in the tumor. 
At the neck, glucose showed slight increase 
but lactate did not.
Fat remained almost constant in time,
with larger amount around the neck where more fat is located.

\section{Discussion}
In order to validate our method, we first performed the phantom experiments.
Figure~\ref{fig:dynamics_Phantom} shows that the spatio-temporal 
distributions of glucose, lactate, and fat are appropriately 
reconstructed by our method, 
and they are consistent with the settings of the respective experiments. 
Figure~\ref{fig:temporal_Phantom} shows the reconstructed 
temporal profiles of the amounts of the substances  
at spatial pixels where the reconstructed profiles attained 
the largest values. 
In Experiments 1--3, 
the glucose increased and then remained almost constant. 
The time at which it reached the end of the increase corresponded approximately 
to the time at which the instillation of the solution actually ended 
($\bigtriangledown$ in Figure~\ref{fig:temporal_Phantom}). 
Lactate continued to increase until the end of the experiment, 
which was consistent with the experimental setting 
that the instillation completed at the end of the measurement.
In Figure~\ref{fig:temporal_Phantom} (c), 
the amount of fat did not fluctuate with time, 
which was consistent with the fact that the amount of fat in a mouse did not 
change in the short period of time.
On the basis of the above, we have succeeded in demonstrating 
the spatio-temporal distributions of the three substances 
with a temporal resolution of four seconds from undersampled MRSI data.

In Figure~\ref{fig:temporal_Phantom}, 
we observe that in Experiments 1-4 and 2, 
the amounts of the substances were reconstructed to be zero for several minutes 
after the start of the measurement, 
even though the instillation of the solution started at the beginning of the measurement.
Since our reconstruction method employs $\ell_1$-norm regularization, 
small values tend to be estimated as zero.
The fact that the amounts of the substances were estimated to be zero continuously after the beginning 
can be attributed to the effect of the $\ell_1$-norm regularization term in $\bm{x}$.
In addition, for glucose and lactate in Experiments 1-1, 1-2, 1-3 and 3, 
the actual amount of substance is zero at the beginning of measurement, 
but is reconstructed to be a constant non-zero value for several minutes after the beginning.
This could be due to the effect of the $\ell_1$-norm regularization term 
of $x_{m+1}-x_{m}$, which encourages the estimates to satisfy $x_{m+1}-x_{m}=0$.
Our reconstruction method interpolated the missing measurement data 
by performing $\ell_1$-norm regularization based on the prior knowledge 
in the framework of compressed sensing.
Although the effect of $\ell_1$-norm regularization can be the limitation of our method, 
the magnitude of the effect of $\ell_1$-norm regularization can be changed 
by adjusting the regularization parameter.
The regularization parameters were determined by CV in this paper, 
but it may be better to determine them 
according to the variation of the amount of substance to be visualized.
In the CV of this paper, the whole measured MRSI data was equally divided into two data sets, 
and the regularization parameters were chosen 
to reduce the error in estimating the data in the data set
not used in the estimation. 
In Experiment 1, the time interval in which the amount of glucose was high 
was the longest in Experiment 1-1 and the shortest in Experiment 1-4. 
Therefore, it is possible that the CV error tended to be smaller 
when the estimated MRSI data was fitted to the measured MRSI data 
in the time interval with high glucose content, especially in Experiment 1-1.
The estimation results may be more consistent with the experimental settings 
if the regularization parameters are chosen separately 
for the time periods with large and small changes in amount.

In the results of the animal experiment, the estimated spatio-temporal distributions of glucose, lactate, and fat,
as presented in Figs.~\ref{fig:dynamics_animal} and \ref{fig:temporal_animal},
are consistent with our expectations:
The initial increase of glucose at the injection site,
followed by its spread over the whole body and decay,
reflects normal physiological processes.
The constancy of fat is explained by the endogenous natural abundance of $^{13}$C.
The elevation of lactate level within the tumor 
following the rise of glucose is ascribed to the Warburg effect.
With regard to the Warburg effect, it is said that glucose 
begins to change into lactate in a few seconds, 
and the temporal change of glucose and lactate could not be 
observed by 2D spectral MRSI with full sampling due to the prolonged acquisition time. 
Our method allows us to reconstruct the spatio-temporal distribution of the substances 
with a temporal resolution of four seconds, 
which enables us to observe the Warburg effect in vivo noninvasively.

One can observe, in Fig.~\ref{fig:temporal_animal}, step-like structure 
in the first half of the temporal profile for glucose at the injection site,
as well as those for glucose and lactate inside the tumor. 
It seems that these parts of the temporal profile include some stair-step artifacts,
because one generally expects smooth temporal changes in the distributions, 
except for the rapid increase of glucose due to its injection.
Similar step-like structures can be observed in the results
of the phantom experiments as well (Fig.~\ref{fig:temporal_Phantom}).
It is considered that this step-like artifact is not due to noise
in measurement but is caused by the regularization. 
We observed dependence of the stair-step artifacts on the regularization parameters $\lambda_\mathrm{W1}$ and $\lambda_\mathrm{W2}$ (results not shown), 
suggesting that they were indeed artifacts: 
Smaller $\lambda_\mathrm{W1}$ or larger $\lambda_\mathrm{W2}$ 
yielded the estimated temporal profiles smoother.
Although the used values of the regularization parameters in our experiments achieved the smallest RMSE in CV, 
there were other sets of the regularization parameters
with RMSEs which were not much different from the one with the smallest RMSE. 
For example, the parameter values 
$(\lambda_{\mathrm{x}},\lambda_{\mathrm{W1}},\lambda_{\mathrm{W2}})
=(10^0,10^2,10^2)$ 
gave rise to a relatively small RMSE (Fig.~\ref{fig:CV} (b)). 
The resulting temporal profiles became smoother without any noticeable stair-step artifacts, but the rapid increase of glucose at the beginning disappeared.
Even if some stair-step artifacts are observed, 
if we do not strongly regularize the sparsity of the temporal variation of the substance distribution, 
we will not be able to observe the rapid changes.
Therefore, the parameter value $(\lambda_{\mathrm{x}},\lambda_{\mathrm{W1}},\lambda_{\mathrm{W2}})
=(10^0,10^3,10^{-1})$ used in our experiment was suitable for our measurement data where we want to detect the amount of substances that changes quickly in time. 
The values of the regularization parameters should be chosen according to the temporal variation of the substance distribution.

In our method, we assume that the base spectra for the substances of
interest are known.
In our experiments, the MRS measurements to prepare the base spectra
were performed simultaneously with the MRSI measurement. 
If it is not the case, then an alternative approach would be to estimate them jointly
with the estimation of $\bm{x}$,
via regarding~\eqref{eq:theta} as defining a PSF model,
where the base spectra $\Theta_\mathrm{B}$ are assumed independent of
positions or time of appearance of the substances.
This assumption is essentially equivalent to that in \citet{Lam2014,Ma2016},
and thus efficiently dealt with
via low-rank matrix decomposition methods \citep{Liang2007,Haldar2010}. 
As another way to prepare the base spectra of substances, 
it would be created from the already acquired MRSI data. 
Using all acquired MRSI signals 
without considering temporal or spatial resolution, 
a combined spectrum of all substances existing in the imaging area 
is reconstructed. 
Characteristic frequencies of spectral peaks
are known for many substances.
By analyzing the reconstructed spectrum, 
it is possible to determine what substances exist in the object 
and to extract the spectrum of each substance. 
By using the extracted spectra of the substances as base spectra of the substances, 
the reconstruction can be performed by our method without additional measurement experiments.
If additional measurement experiments are available,
one may alternatively perform MRS measurements on a stable phantom with known substance concentrations in order to prepare the base spectra. 
The approach with measurements for the base spectra in advance opens a way 
to perform quantitative analysis on spatio-temporal substance distributions from MRSI data.
This will be future work on development of MRSI framework.

The advantage of this method is 
that it can reconstruct the spatio-temporal distributions of multiple substances simultaneously on a time scale that is much shorter than the time scale of full MRSI sampling. 
Another advantage is that it does not require training with a large amount of data in advance and can easily reconstruct the substance distributions using only the acquired MRSI signal.
However, as a limitation, this method cannot reconstruct whole MRSI spectra 
that include various spectral peaks other than those of the substances of interest, 
because the method reconstructs the substance distributions 
using the base spectra that include the peaks of the substances of interest. 
One idea to break through this limitation is 
not to decide on specific substances as the targets of estimation, 
but to estimate the whole spectra 
by combining spectra estimated separately 
for water, lipid, metabolite, and macromolecules, as is done in \citet{Lam2020}. 
Another idea is to use a large number of spectra with various peaks as base spectra instead of spectral peaks of specific substances, 
to reconstruct the substance distributions without determining the substance of interest at the reconstruction stage, 
and to synthesize the entire spectrum to obtain the MRSI spectrum. 
However, this method would have a very large number of parameters to be estimated, so it may be necessary to collect some information in advance by some technique such as machine learning with a large amount of data.

In the experimental setting of this study, 
we used a 2D spectral and 2D spatial MRSI sequence. 
In the Fourier space of $\mathcal{F}(Q)\times\mathcal{F}(R)$, 
$\mathcal{F}(Q_2)$ is the readout dimension in which the MRSI signal is acquired at one time. 
The readout is performed per point in the remaining three dimensions 
of $\mathcal{F}(Q_1)\times\mathcal{F}(R_1)\times\mathcal{F}(R_2)$. 
The required time for a single readout per point is four seconds. 
The resolution of the spectrum was 256 (in the readout direction) $\times$ 32, 
and the resolution of the space was 8 $\times$ 8 or 8 $\times$ 16. 
In order to acquire complete information about the 2D spectrum in 2D space, 
32 times sampling in the $\mathcal{F}(Q_1)$ direction and 8 $\times$ 8 or 8 $\times$ 16 times sampling in the $\mathcal{F}(R)$ direction are required. 
In other words, to obtain a single 2D image where each spatial pixel has 2D spectral information, 4 s $\times$ 32 $\times$ 8 $\times$ 8 (or 16) $=$ 2.3 hours (or 4.5 hours) is required. 
Therefore, reconstruction from a fully sampled signal only produce a movie with one time frame of 2.3 hours (or 4.5 hours).
In this study, we propose a method to randomly acquire signals in the three dimensions of $\mathcal{F}(Q_1)\times\mathcal{F}(R_1)\times\mathcal{F}(R_2)$ 
and to reconstruct the substance distributions for each time frame divided by an arbitrary time width from the time-series of the MRSI signals. 
For the spectral dimension, 
we have reduced the number of elements to be reconstructed 
by using the base spectrum of each substance. 
By using compressed sensing on the spatial dimensions $R$, 
we have succeeded in obtaining a movie with a time resolution of four seconds. 
The time width of the time frame can be determined arbitrarily. 
In other words, the number of acquired signals used in one time frame can be arbitrarily determined. 
The four seconds is a time elapsed during one readout. 
It is computationally possible to assign fewer signals 
to one time frame than the number of signals for one readout.
However, since four-step phase cycling was employed in this experimental condition, 
signals acquired through four times signal acquisitions in different phases were combined 
assuming that the substance distributions did not change, during this time.
Therefore, it is unreasonable to estimate the substance distribution from the signal in the shorter time than the readout time. 
For this reason, four seconds is considered to be the lower limit of the time resolution under the conditions of this experiment, 
but it could be shortened further by using other MRSI acquisition sequences, 
such as ultrafast MRSI imaging with signal enhancement by hyperpolarization \citep{Muller2020}.

As mentioned in the previous paragraph, 
in this study, a temporal resolution of four seconds was achieved by reconstructing the substance distribution image of one time frame per MRSI signal obtained as a readout at a single point in the three dimensions of $\mathcal{F}(Q_1)\times\mathcal{F}(R_1)\times\mathcal{F}(R_2)$.
Since we compressed the spectral dimension using the base spectra of target substances, 
the required full sampling points depend on the number of target substances.
The CS acceleration factor (the under-sampling ratio for CS compared to full sampling to reconstruct a spatial distribution of substances) 
is calculated to be 3 (number of target substances) $\times$ 8 $\times$ 16 $=$ 384 in the animal experiments, 
which is an extreme acceleration. 
However, this method does not reconstruct the whole MRSI spectra, nor does it reconstruct the substance distributions from only the MRSI signal assigned to a single time frame.
Therefore, it is not fair to compare the temporal resolution with those in existing reports that reconstruct the spectrum of a single image. 
To the best of our knowledge, 
there are no reports on methods for estimating spatio-temporal distributions of substances 
from MRSI signals with the same strategy as this study, 
so we cannot compare them directly. 
In \citet{Klauser2021}, they have achieved CS acceleration factors of up to 6.5 in 1D-spectral 2D-spatial MRSI.
A study has reported that the acquisition time was 0.9 minutes for a spatial matrix of size 17 $\times$ 17 $\times$ 5 in 1D MRSI \citep{Li2021}. 
On the other hand, 
when imaging is performed using hyperpolarization techniques, the MRSI signal is amplified tens of thousands of times, so the signal can be acquired in a very short measurement time and the temporal resolution can be very high. 
Therefore, it is also not fair to compare the time resolution with experiments that do not use such signal amplification during imaging, as in this study. 
In \citet{Muller2020}, they have reported that it took only 8.2 seconds to measure 1D MRSI in 3D space.

\section{Conclusion}
We have reconstructed the spatio-temporal distributions 
of substances with high temporal resolution from the undersampled 2D spectral MRSI data using compressed sensing.
Our method utilized 
the acquired MRSI signal and the base spectra of substances of interest. 
We exploited the base spectra as well as the temporal change of the distribution of the substances as the prior knowledge for the compressed sensing. 
Our method 
was successful in reconstructing spatio-temporal distribution of multiple substances only from undersampled MRSI data with base spectra of substances of interest. 
Using our method, it will be possible to observe the spatio-temporal dynamics of substances with high temporal resolution from existing MRSI data.
Indeed in our experiments the temporal resolution was four seconds in spite of long acquisition time for full sampling.
Our method allows us to observe phenomena where the spatio-temporal distributions of substances change in a short time in vivo, such as metabolic activities and medical efficacy, and has the potential to capture unknown changes of substances  distributed in vivo.
In future work, it is expected that the reconstruction will be performed by preparing a large number of base spectra of substances without specifying the target substances, which will lead to the discovery of substances whose existence in vivo is unknown.

\begin{acknowledgments}
This work was supported in part by the MEXT/JSPS KAKENHI 
[grant numbers JP25120008, JP16H02878, JP16K16407, JP19K20709, JP20J40290, and JP22K12840].
\end{acknowledgments}

\bibliographystyle{apa.bst}
\bibliography{MRM180926bib_CleanVersion}

\end{document}